\documentclass[epj]{svjour}
\usepackage{graphicx}
 \usepackage{amsmath}
 \usepackage{amsfonts}
 \usepackage{amssymb}
 \usepackage{pstricks}
 \usepackage{psfrag}
 \usepackage{epsfig}
 \usepackage[ansinew]{inputenc}

 \renewcommand{\vec}[1]{\mathbf{#1}}
 
 \newcommand{\mat}[1]{\underline{\underline{#1}}}	
 \newcommand{\matij}[2]{{#1}_{#2}}			
 \newcommand{\Wi}{\text{W}}				
 \newcommand{\Eq}[1]{Eq.~(\ref{#1})}			
 \newcommand{\se}[1]{Sect.~\ref{#1}}			
 \newcommand{\fig}[1]{Fig.~\ref{#1}}			

\begin{document}

\title{Dumbbells in suspension: A numerical study on their dynamics and
 shear viscosity
}

\author{Johannes Greber$^{1}$, Jochen Bammert$^{1}$, Philippe Peyla$^{2}$, Walter Zimmermann$^{1}$}

\institute{$^1$ Theoretische Physik I, Universit\"at Bayreuth, 95440 Bayreuth, Germany \\
$^2$ Laboratoire de Physique Interdisciplinaire, UMR 5588, Universit\'e Joseph-Fourier, 38402 Saint Martin d'Heres, France}

\date{Received: April 26, 2013 / Revised version: (date)}

\abstract{
The dynamics of elastic dumbbells in linear shear flow is investigated 
 by fluid particle dynamics simulations at small Reynolds numbers. 
The positive contribution of a single dumbbell
to the effective shear viscosity 
is determined via the extra stress exerted at the boundaries of the
shear cell and the difference  to the contributions 
obtained via the Kramers-Kirkwood formula are described. 
 For a small  Weissenberg number 
and when the mean dumbbell length becomes larger than the mean next-neighbor distance,
  the contribution of interacting dumbbells to the mean shear viscosity
exceeds significantly the  contribution of unconnected beads 
occupying the same volume fraction.
}

\maketitle

\section{Introduction}\label{sec: intro}

Particles in a fluid change the viscosity and 
cause  a variety of interesting as well as
astonishing flow phenomena \cite{Larson:99,Bird:87,Steinberg:2000.1}.
The shear viscosity of a suspension 
increases in the diluted regime linearly  with the 
particle concentration  
 \cite{Einstein:1906.1,Einstein:1911.1} and quadratic effects 
come into play with the importance of 
 particle-particle interactions \cite{Batchelor:1972.1,Felderhof:1988.1}. 
Soft particles, such as polymers, may change with their
extra dynamical degrees of freedom the macroscopic flow
properties of a fluid significantly, especially when the product 
of the particle relaxation-time and
the local shear rate, the so called Weissenberg number,
reaches values of the order one. In this regime,  
but still at small values of the particle Reynolds number,
polymers may exhibit a rich dynamics and
  may give rise  to  phenomena  such as shear thinning  \cite{Larson:99,Bird:87}, 
tumbling 
\cite{Chu:1999.1,Chu:2005.3,Chu:2005.2,Steinberg:2006.3,Turitsyn:2005.1,Gompper:2011.1},
 elastic turbulence \cite{Steinberg:2000.1} or polymer induced, efficient fluid mixing \cite{Steinberg:2001.2}.
Vesicles in linear shear flows are deformed  as well, 
leading to various kinds of nonlinear transitions  between dynamical
states \cite{Misbah:2006.1,Steinberg:2006.1}, a topic  reviewed recently in 
Refs.~\cite{Misbah:2009.1,Steinberg:2011.1}.

Flows of suspensions of 
polymers or vesicles are often modeled 
 by using appropriate constitutive equations for the
stress tensor ${\mat{\sigma}}$ in generalized 
Navier-Stokes equations \cite{Bird:87}. Such models are quite
common in the range
of  small values  of the Weissenberg number, corresponding 
to small particle deformations,  and in
the  regime of small particle concentrations,
where particle-particle interactions are still weak.
The  interactions among many deformable particles in
flow may be intricate, especially the nonlinear, long-range
hydrodynamic interaction as well as effects of walls may 
induce  a complex dynamics. 
Which degrees of freedom are relevant
in  macroscopic modeling?
For this purpose investigations of the dynamics of bead-spring models
in flow by computer simulations
\cite{Fraenkel:1952.1,Warner:1972.1,Bird:1975.1,Leal:1975.1,Swaroop:2006.1,Owens:2010.1,Winkler:2009.1}
 are a powerful tool. Appropriate simulations enable an estimate, for instance,
 of the relative importance of the dynamics of single soft particles in flow and
 the particle-particle interactions
\cite{Fraenkel:1952.1,Warner:1972.1,Bird:1975.1,Leal:1975.1,Swaroop:2006.1,Owens:2010.1,Winkler:2009.1}.
Well known computational schemes 
 are Brownian dynamics simulations \cite{dePablo:2003.1,Prakash:2004.1},
multi-particle-collision dynamics \cite{Kapral:2000.1,Gompper:2008.1} or
lattice Boltzmann methods \cite{Ladd:2010.1}.

Here  we investigate the dynamics of a suspension
of elastic dumbbells by three dimensional simulations  
and calculate their contribution to  the shear viscosity 
of the suspension.  For this purpose we use fluid particle 
dynamics (FPD), which was developed by Tanaka and Araki \cite{Tanaka:2000.1}
and which is  summarized in \se{sec: model}. 
It is based on a continuum description of
suspended particles  in an incompressible 
fluid, where  the viscosity of the solvent is enhanced
at the positions of the particles.
This approach has been  applied successfully to 
explorations of the dynamics of colloidal particles in 
complex fluids \cite{Tanaka:2008.1}, the polymer 
coil-globule transition \cite{Tanaka:2009.1}, the effect of confinement on the rheology 
of a suspension of spheres \cite{Peyla:2008.1}, and the effective viscosity of
micro-swimmer suspensions \cite{Peyla:2010.1}.

The tumbling dynamics of a single
dumbbell in the shear plane   
 and the  contribution of a dumbbell 
to the effective shear viscosity  is determined
in a direct manner via the applied shear stress 
at the walls of a shear cell in \se{sec: singdumb}.
The difference between these direct contributions to the shear stress
 and the dumbbell contribution 
via the Kramers-Kirkwood approach for deformable dumbbells \cite{Bird:87,DoiEd:86,Winkler:2013.1} 
are also described.
The  effects of many interacting dumbbells in a shear flow
is investigated in 
\se{sec: susp}, where we present a significant enhancement 
of the effective viscosity
in the case of an increasing importance of the dumbbell-dumbbell correlation.

\section{Fluid Particle Dynamics and effective viscosity}\label{sec: model} 

In fluid particle dynamics simulations a suspension 
is described by an incompressible single component fluid 
with a spatially varying viscosity  \cite{Tanaka:2000.1},
\begin{align}
\label{eq_etar}
 \eta(\vec{r})= \eta_s + \frac{\eta_p-\eta_s}{2}\sum_{i=1}^N \left[ 1+ 
 \tanh \left( \frac{\nu-|\vec{r}-\vec{r}_i|}{\xi} \right) \right]\,,
\end{align}
where the viscosity is considerably enhanced 
at the particle positions ${\bf r}_i$ ($i=1,  \ldots, N$)
 to about  $\eta_p=100\eta_s$. 
The parameter $\nu$ describes the 
range of the local viscosity enhancement 
around ${\bf r}_i$ and $\xi$ the width of the 
transition regime between the two viscosity levels $\eta_s$ and $\eta_p$. 
Throughout of this work we 
 choose $\nu=2$ and $\xi=0.5$, which corresponds to
an effective particle radius $a=3$.

The dynamics of the velocity field $\vec{u}(\vec{r},t)$
 of an incompressible fluid of density $\rho=const.$ ($\nabla \cdot {\vec u}=0$)
is described by  the time-dependent equation
\begin{align}
 \label{eq_ustokes}
\rho \left[\partial_t \vec{u}   +(\vec{u}\cdot \nabla) \vec{u} \right]
=\nabla \cdot \mat{\sigma} \,\,,
\end{align}
where  the stress tensor 
\begin{align}
 \label{eq_gs}
 {\mat{\sigma}}= -p{\mat{I}}+\eta \left(\nabla\vec{u}+(\nabla\vec{u})^T \right)\,.
\end{align}
includes the spatially varying viscosity field $\eta(\vec{r})$
and   the pressure field $p({\bf r})$ with  the unity matrix  ${\mat{I}}$.
\Eq{eq_ustokes} is solved on a three dimensional grid (Marker and Cells) 
with a mesh of length one, by
using a projection method \cite{Peyret:1990}. The
 resulting velocity field,  averaged over the range
of the enhanced viscosity, is used to update the particle's 
position, as described in more detail in  Refs. \cite{Tanaka:2000.1,Peyla:2007.1}.

\begin{figure}[ht]
  \begin{center}
  \includegraphics[width=0.8\columnwidth]{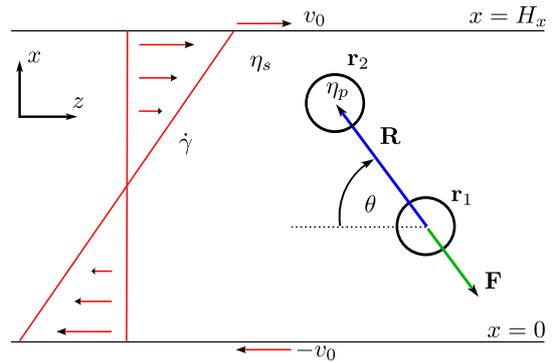}
  \end{center}
  \caption{In FPD suspended particles are modeled by an enhanced 
viscosity  $\eta_p=100\eta_s$ around their centers at $\vec{r}_{i}$. 
A dumbbell consists of two beads connected by a  spring along the 
connection vector $\vec{R}=\vec{r}_{2}- \vec{r}_{1}$.
Moving the upper and lower plate of the flow channel with constant velocity 
$v_0$, we create a linear shear flow with shear rate $\dot \gamma=2v_0/H_x$. 
}
\label{fig1}
\end{figure}

The two beads of a dumbbell are
connected by a linear spring with an equilibrium
length  $R_0$ of the connection vector 
$\vec{R}=\vec{r}_2-\vec{r}_1$, cf. \fig{fig1},
and the spring constant $k$. 
 The resulting  spring force  
\begin{align}
 \label{eq_springforce}
 \vec{F}(\vec{R})=f(R) \frac{\vec{R}}{R}\,
\end{align}
is proportional the elongation $R(t)-R_0$ 
with $f(R(t)) =k[R(t)-R_0]$ and acts along the connection vector ${\bf R(t)}$.
 For the sake of simplicity we use harmonic springs,
but for FENE springs, as described in Ref.~\cite{Warner:1972.1},
qualitatively similar results are obtained.
$\theta \in [0,\pi]$ is the angle enclosed 
by $\vec{R}$ and the undistorted flow direction $\hat{\vec{z}}$.
In order to exclude overlap between dumbbell beads,
we use a short-range repulsive potential $\phi_{LJ}= \alpha/r^{12}-\beta/r^6$ 
for each bead with a cut-off length of $2a$.

The dumbbells are confined in a rectangular flow channel of
size $H_x \times H_y \times H_z$ with $H_x=H_y=60$ and $H_z=100$, if not
stated otherwise.
We impose no-slip boundary conditions at the walls in $x$ and
$y$ direction
and periodic boundary conditions along the $z$ direction.
By moving the upper plate at $x=H_x$ with a velocity $v_0\hat{\vec{z}}$ and 
the lower plate at $x=0$ with $-v_0\hat{\vec{z}}$, as indicated 
in \fig{fig1}, we create a
linear shear flow of  shear rate
\begin{align}
 \label{eq_dotgamma}
\dot{\gamma}=\frac{2v_0}{H_x}\,.
\end{align}
 For a dumbbell a typical time scale $\tau$ may
be introduced in terms of the 
Stokes friction $\zeta=6\pi\eta_s a$ as well as 
the dimensionless Weissenberg number $\Wi$:
\begin{eqnarray}
 \tau=\frac{\zeta}{k}\qquad \mbox{and}\qquad \Wi=\dot \gamma \tau\,.
\end{eqnarray}

The effective viscosity of a suspension at constant shear stress
is determined in our system by the $xz$ component of the stress
tensor $\mat{\sigma}$ averaged over the whole surface of the moving 
boundaries:
\begin{align}
 \label{eq_etaeffshear}
 \eta_{eff}(t)=\frac{ \matij{\bar\sigma}{xz}(x=H_x) +
\matij{\bar \sigma}{xz}(x=0)}{2\dot{\gamma}}\,\,.
\end{align}
According to the tumbling motion of  dumbbells
 $\eta_{eff}(t)$ is a function of time
and its time average is denoted by 
\begin{align}
\bar{\eta}_{eff}=\langle {\eta}_{eff}(t)\rangle\,.
\end{align}
The effective averaged viscosity of a homogeneous 
suspension of independent spheres can be 
represented in terms of a virial expansion with respect to the volume 
fraction $\Phi$ occupied by the particles in suspension:
\begin{align}
  \label{eq_ebg}
  \bar{\eta}_{eff}(\Phi)=\eta_s\left(1+w_1\Phi +w_2\Phi^2+\mathcal{O}\left(\Phi^3\right)\right) \,.
\end{align}
In the dilute regime, where hydrodynamic particle-particle interactions become negligible,
Einstein determined the first coefficient: $w_1=2.5$ \cite{Einstein:1906.1,Einstein:1911.1}.
The second order coefficient $w_2=5.2$
was determined by Batchelor and Green \cite{Batchelor:1972.1}, 
which has later been evaluated including the high frequency regime as well
by Felderhof and Cichocki to $w_2=5.0$ \cite{Felderhof:1988.1}.
In FPD simulations it was shown that both coefficients $w_1$ and $w_2$ are  changed 
in the case of strong confinement of a suspension of
spheres   \cite{Peyla:2010.1}. How the
 shape of  suspended objects 
influences  the viscosity have been also investigated \cite{Brenner:1981}.

Two useful quantities for  a characterization 
of  the dumbbell dynamics and the shear viscosity 
are  the relative deformation
\begin{align}
\label{epsdef}
 \varepsilon(t)=\frac{R(t)-R_0}{R_0}\,,
\end{align}
and the spatially averaged relative viscosity change induced by
the dumbbells,
\begin{align}
 \label{eq_releta}
 \Delta\eta(t)=\frac{\eta_{eff}(t)-\eta_s}{\eta_s}\,.
\end{align}
In simulations of colloidal suspensions the contributions of
particles to the shear viscosity are often determined indirectly
via the  Kramers-Kirkwood formula  of the stress tensor
 \cite{Bird:87,DoiEd:86},
\begin{align}
 \label{eq_lambda}
  \bar {\mat{\lambda}} = - \langle \vec{R}(t) \otimes \vec{F}(t) \rangle \,,
\end{align}
whereby this formula is derived under the assumption of 
 point like beads at the ends of  dumbbells, i. e.  with 
the bead radii much smaller than the bead distance, $a \ll R_0$. 
The time dependence of the connection vectors   $\vec{R}_i(t)$ 
of  dumbbells  and the spring forces
${\bf F}_i(t)$  in this formula may be obtained from 
numerical simulations of the dumbbell dynamics.

In this work we evaluate the formula (\ref{eq_lambda}) 
only for a single dumbbell in order
to estimate the differences between the dumbbell contribution
to the viscosity, when the dumbbell is composed 
of point like beads, and a direct determination
of the dumbbell contribution to the shear
viscosity change  in Eq.~(\ref{eq_releta}).
For this purpose we take the
time dependence of ${\bf R}(t)$ obtained  from simulations
and  calculate  the time dependence of $\bar {\mat{\lambda}}(t)$ 
and compare it with   Eq.~(\ref{eq_releta}).

\section{Dynamics of a single dumbbell \label{sec: singdumb}}
In this section, the  rotation of a dumbbell in  the shear plane
 is investigated, where its contribution to the   relative shear viscosity
change $\Delta \eta(t)$ takes its largest values. This is compared
with the contribution  of two independent beads to 
$\Delta \eta(t)$ as well as
with the time dependence of the  dumbbell contribution 
to the stress tensor,  as calculated in terms of the dumbbell conformation
via the  Kramers-Kirkwood formula given by Eq.~(\ref{eq_lambda}).

\begin{figure}
  \begin{center}
  \includegraphics[width=1.0\columnwidth]{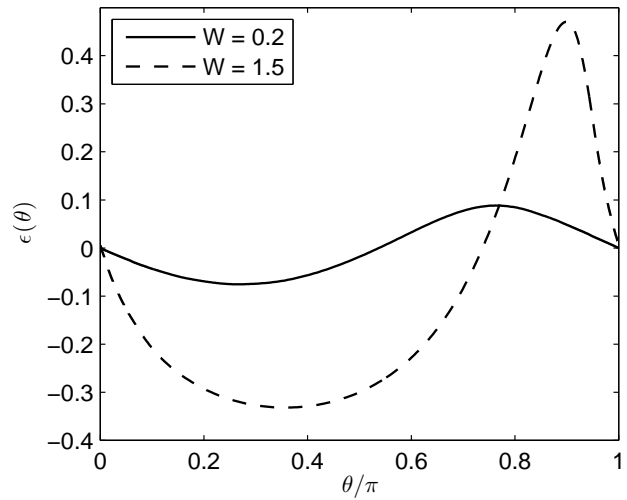}
  \end{center}
  \caption{The relative dumbbell deformation
$\varepsilon(\theta)$ [cf.  Eq.~(\ref{epsdef})]
is shown during one half turn in the  shear plane 
for  $R_0=10$  and 
 two different
values of the Weissenberg number: $\Wi=0.2$ and $W=1.5$.
}
\label{fig2}
\end{figure}

With increasing values of the Weissenberg number, $ \Wi$,
the duration of a dumbbell rotation 
in the shear plane decreases 
and the temporal dumbbell deformation increases.
The compression (stretch)
of a dumbbell spring increases during the
first (second) quarter of 
a dumbbell turn,
as indicated  in \fig{fig2} by the relative extension  $\varepsilon(\theta)$ 
  for  $\Wi=0.2$ and $\Wi=1.5$. In the
 range of $\Wi   \ll 1$ the dumbbell
deformation adapts nearly instantaneously to the
balance between the spring and friction force.
Accordingly, the dumbbell deformation $\varepsilon(\theta)$
vanishes at about $\theta \sim \pi/2$, where for a short moment
during a dumbbell turn either compressing or extending -
frictional forces become rather small.

This is different in the  range of larger values of $\Wi$,
where dumbbell turns are faster with respect to
the relaxation time. In this range 
the dumbbell relaxation follows with a finite delay the temporal difference
between the spring and  the friction forces acting on the dumbbell beads.  
For the Weissenberg number  $\Wi=1.5$ a dumbbell 
is strongly compressed  by  viscous friction forces 
during the first quarter of a turn, which is followed by 
a rather quick dumbbell rotation on the scale of the
scale of the relaxation time, so that the dumbbell spring is  still 
not relaxed when the dumbbell orientation passes 
the angle  $\theta =\pi/2$. In this case
the relative dumbbell deformation $\varepsilon(\theta)$ 
becomes zero at a delayed angle $\theta_0>\pi/2$.
The phase shift $\theta_0-\pi/2>0$ increases with the Weissenberg number 
 $\Wi$ and  causes an asymmetric $\theta$ dependence  of 
$\varepsilon(\theta)$ with respect to $\pi/2 <\theta_0 <\pi$,
 as shown in \fig{fig2}. 

\begin{figure}
  \begin{center}
  \includegraphics[width=1.0\columnwidth]{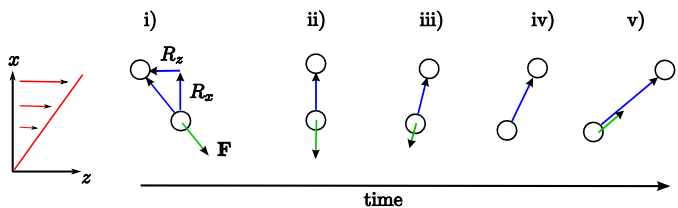}
  \end{center}
  \caption{The sketch of the tumbling motion at a high 
 Weissenberg number  illustrates 
  the connection between dumbbell dynamics and $\matij{\lambda}{xz}(t)$.}
  \label{fig3}
\end{figure}

A complementary view on the angle dependence of the relative deformation 
$\varepsilon(\theta)$ provides \fig{fig3}. This figure
shows  the connection vector ${\bf R}$ (blue)
at different values of the dumbbell orientation angle $\theta$  as well as  the
spring force $\vec{F}$ (green) acting on one bead,
whereby  $-{\bf F}$ acts on the opposite bead.  The figure 
indicates that the force ${\bf F}$, being either parallel
or anti-parallel  to $\vec{R}$, vanishes and changes 
its sign at the angle $\theta_0 > \pi/2$, where the
relative extension $\varepsilon(\theta)$ passes zero
in  \fig{fig2}.

\begin{figure}
  \begin{center}
  \includegraphics[width=1.0\columnwidth]{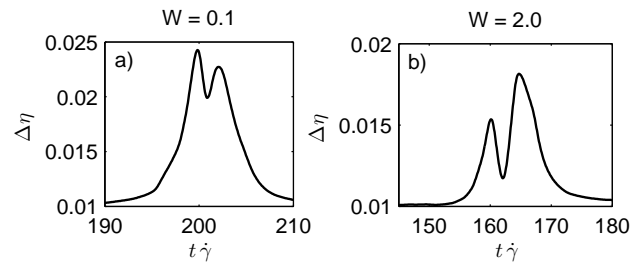}
  \end{center}
\vspace{-5mm}
  \caption{
Part a) and b) show the time dependence of the
relative viscosity chance $\Delta \eta(t)$ during 
one half turn of a dumbbell at two different values of the Weissenberg number:
 $\Wi=0.1$ (left) and $\Wi=2.0$ (right).
}
  \label{fig4}
\end{figure}
\begin{figure}
  \begin{center}
 \includegraphics[width=1.0\columnwidth]{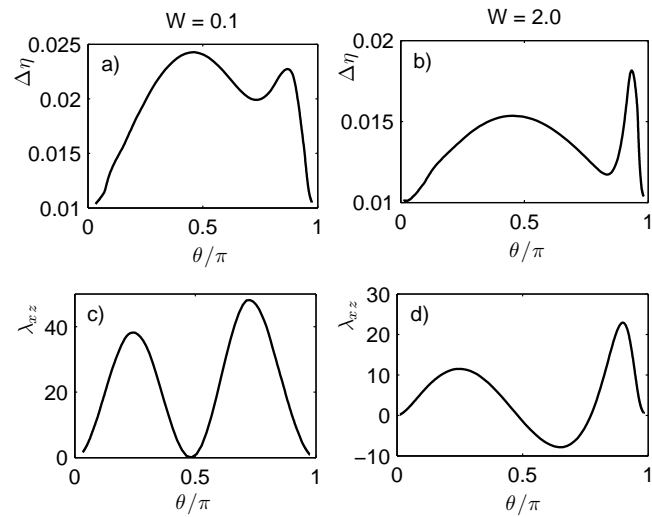}
  \end{center}
  \caption{Part a) and b) show the relative viscosity chance $\Delta \eta(\theta)$ during 
one half turn of a dumbbell at two different values of the Weissenberg number:
 $\Wi=0.1$ (left) and $\Wi=2.0$ (right).
Parts c) and d) show  $\lambda_{xy}(\theta)$
for the same parameters.
}
  \label{fig5}
\end{figure}

The contribution of a dumbbell 
to the relative shear-viscosity change $\Delta \eta(t)$
varies as a function of time and the orientational angle $\theta$
as illustrated by numerical data 
in   \fig{fig4} and 
in the upper part of \fig{fig5}  for two different values of $\Wi$.

The contribution of deformable particles in a fluid 
to the shear viscosity is often calculated
via the stress tensor $\lambda_{ij}$ \cite{Bird:87}, which may be
calculated in terms of deformation data obtained from simulations.
For a dumbbell the matrix element  
 $\lambda_{xz}(t)$ in  Eq.~(\ref{eq_lambda})
takes the following form:
\begin{align}
 \label{eq_rfxz}
 \matij{\lambda}{xz}(t)=-f(R)\frac{R}{2}\sin(2\theta)\,.
\end{align}
It depends on the dumbbell extension $R(t)-R_0$ via $f(R)$
and on the orientational angle $\theta$.
For small values of $\Wi$ the two functions $\sin(2\theta)$
and $R(t)-R_0$ pass  zero near the perpendicular
dumbbell orientation, i.~e.  for $\theta\sim \pi/2$,
and  therefore   $\lambda_{xz}(\theta\sim \pi/2)$
vanishes as  in \fig{fig5}c).
For larger values of $\Wi$ the function
$R(\theta)-R_0$ passes zero at
an angle $\theta_0$  larger
than $\pi/2$, cf. \fig{fig2},
and $\lambda_{xz}(\theta)$ becomes negative
between both zeros, 
as can be seen in \fig{fig5}d).
On the other hand the directly determined
relative viscosity change $\Delta \eta$ is always positive
in \fig{fig5}a) and \fig{fig5}b) -  as expected. The fact, that  $\lambda_{xz}$ becomes
negative in an intermediate range, shows the limitations
of the  Kramers-Kirkwood-formula  for a determination of
the contribution of deformable
particles to the stress tensor of a suspension
and therefore  to the shear viscosity 
of a suspension of dumbbells or
other deformable particles.

\begin{figure}
 \begin{center}
  \includegraphics[width=0.52\columnwidth]{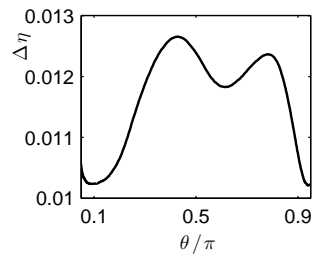}
\end{center}

  \caption{The contribution of two unconnected beads 
to the relative viscosity chance $\Delta \eta(t)$ in 
a shear cell as a function of the  angle $\theta$,
enclosed by the connection vector between two beads and 
the streamlines.}
  \label{fig4a}
\end{figure}

In \fig{fig4}a) and  \fig{fig4}b)
the dumbbell axis is nearly parallel to the flow lines
for small and large values of $\dot  \gamma t$.
In this range the dumbbell contribution  to $\Delta \eta$
is similar to the contribution of two unconnected beads with
its  connection vector parallel to the flow lines,
as indicated in \fig{fig4a}. In both cases the relative shear viscosity 
change is about $\Delta \eta \simeq 0.01$ and therefore
the  contribution to  $\Delta \eta$ of a dumbbell 
with the connection vector parallel to the flow lines
is  essentially caused by the rotating beads of finite diameter
at both ends of a dumbbell.

 The dependence of $\Delta \eta(\theta)$  on 
the angle $\theta$ of the connection vector between two unconnected beads in \fig{fig4a} 
shows  two maxima, similar as  for a dumbbell in \fig{fig5}a) and \fig{fig5}b). 
The  hydrodynamic interaction between the
two rotating beads is in both cases comparable at similar bead distances. 
This interaction  causes the major contribution to
 $\Delta \eta(\theta)$ in \fig{fig4a} for two unconnected beads,
whereas in the case of dumbbells the spring force causes
 modifications,  leading
to the differences between the results shown in \fig{fig4a} at the
one hand and on the other hand in \fig{fig5}a) and \fig{fig5}b). 

In our numerical examples, the maxima of $\Delta \eta$ are
in the range of small values of $\Wi$ 
for dumbbells  roughly by a factor of two larger as 
for unconnected beads - due to the spring force.
 This enhancement  becomes larger by
decreasing the ratio between the bead diameter
and the dumbbell length. It becomes  smaller by
increasing the Weissenberg number $\Wi$. Note, that
$\Wi$ may be enhanced  by increasing the shear rate
or by decreasing the spring constant, i. e. large values of $\Wi$
are closer to the case of unconnected beads.

The reason for a higher contribution 
of a dumbbell to $\Delta \eta$  is as follows. During a dumbbell turn 
the dumbbell spring is most of the
time either compressed or stretched and the forces  involved
during a  dumbbell compression or stretch are
actually exerted by the moving boundaries
of the shear cell via the viscous fluid,
which causes an enhanced contribution to $\Delta \eta(t)$.
Between the ranges where the dumbbell is compressed
or stretched, the dumbbell passes rather quickly
to a vertical orientation,
as indicated by \fig{fig4}a) and \fig{fig4}b).
In this range the spring becomes relaxed 
and leads to a reduction of  $\Delta \eta$
in an intermediate state.

\section{Viscosity of a dumbbell suspension}\label{sec: susp} 

In a suspension of dumbbells, excluded volume  and
the nonlinear hydrodynamic interactions lead  
to a far more complex dynamics of  dumbbells
 than in the case of 
one isolated dumbbell, as discussed in the previous  \se{sec: singdumb}.
This is demonstrated by simulations of $80$ dumbbells 
in a box filled with a Newtonian fluid, where the
dumbbells are initially placed on a lattice 
with the same orientation angle $\theta$.  
\begin{figure}[ht]
\vspace{-2mm}
  \begin{center}
  \includegraphics[width=0.9\columnwidth]{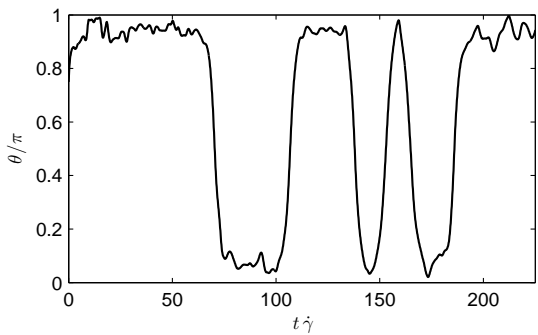}
  \end{center}
 \vspace{-5mm}
  \caption{The temporal behavior of the orientation
 angle $\theta(t)$ of an individual dumbbell
in a suspension of $80$ objects (corresponding to $\Phi= 0.05$)
behaves rather irregular. 
The dumbbell extension was $R_0/a=3.33$ 
and $W=2.0$.}
\label{fig6}
\end{figure}
After a transient regime of redistribution of dumbbells
 in shear flow a  typical temporal 
behavior of the angle $\theta(t)$ of a single dumbbell out of $80$ is
shown in \fig{fig6}.
It confirms that dumbbells are most of their time
with $\theta \sim 0,~ \pi$ nearly
parallel oriented to the undisturbed flow lines parallel to $\hat z$
 and the major part of a dumbbell turn 
takes place during a short period of time.
Compared to the dynamics of a single 
dumbbell in a fluid such turns of  dumbbells
in suspension 
take place rather irregularly in time.

\begin{figure}[ht]
\vspace{-2mm}
  \begin{center}
  \includegraphics[width=0.9\columnwidth]{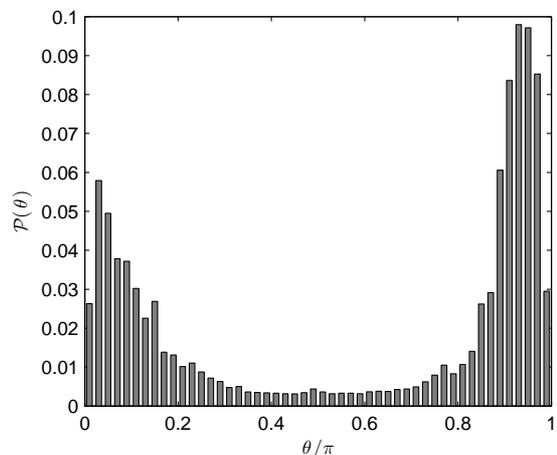}
  \end{center}
 \vspace{-5mm}
  \caption{The time averaged distribution, $\mathcal{P}(\theta)$, 
of the orientation angle $\theta$ 
as obtained for a suspension 
of $80$ dumbbells in shear flow and
Weissenberg number $\Wi=0.21$. }
\label{fig7}
\end{figure}

The stationary orientational distribution 
of  dumbbells and the  effect
of the dumbbell dynamics on the time averaged relative  viscosity change,
\begin{align}
 \Delta \bar{\eta} = \langle \Delta\eta(t) \rangle\,,
\end{align}
are  both measures of the overall mean response
of a suspension to the applied steady shear flow.
The distribution function of the bead positions,
the deformations of springs and the distribution of
the dumbbell orientational angle $\theta$ 
reach in simulations after a transient regime a stationary state, 
independent of the initial condition.
\fig{fig7}  shows the stationary distribution of the dumbbell orientation-angle
$\mathcal{P}(\theta)$, which has
 maxima around $0$ and $\pi$,
because the dumbbell axis between the subsequent half-rotations 
stays most of the time close to $\hat{\vec{z}}$,
cf. \fig{fig6}.

The asymmetry of $\mathcal{P}(\theta)$
and the 
asymmetry of $\varepsilon(\theta)$ in \fig{fig2}
with respect to $\theta=\pi/2$ are
related to each other and  have its origin in 
the hydrodynamic dumbbell-dumbbell interaction  
during the  shear induced rotations of the beads of
the  dumbbells  
and the contribution of the extensional part  of the shear flow.
The hydrodynamic interaction due to bead rotations
is enhanced  during the compressed phase of the dumbbell
and therefore  the dumbbell rotation is quicker
in the range $\theta < \pi/2$, leading to a smaller
probability of the dumbbell orientation 
in this range than in  $\theta > \pi/2$.
This observation is also supported by
the $\theta$-dependence of $\Delta \eta$ 
in \fig{fig5}.

\begin{figure}[ht]
\vspace{-2mm}
  \begin{center}
  \includegraphics[width=0.9\columnwidth]{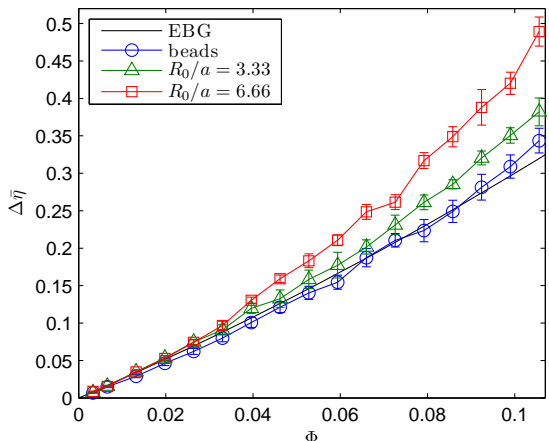}
  \end{center}
 \vspace{-5mm}
  \caption{The averaged relative viscosity $\Delta \bar \eta=\langle \Delta \eta \rangle$ 
is shown for $\Wi=0.33$  as a function of the volume fraction $\Phi$ for a suspension of
spheres (circles) or dumbbells with $R_0/a=3.33$ (triangles) or  
 $R_0/a=6.66$ (squares). The solid line
is according to the analytical result of Einstein, Batchelor and Green
as given by \Eq{eq_ebg},
which deviates from our numerical results beyond $\Phi \simeq 0.08$.
}
\label{fig8}
\end{figure}

According to Eq.~(\ref{eq_ebg}) 
one expects for a diluted suspension of independent spheres
a linear relation between $\Delta\bar{\eta}$ and the volume fraction
$\Phi$. We  tested this in simulations 
by varying    the number of beads  in the shear cell  from $10$ up to $320$, which
corresponds to a variation of the bead-volume fraction
between $\Phi=0.0066$ and $\Phi=0.1056$.
The simulation data obtained for the averaged relative viscosity change $\Delta \bar \eta$
are  given in \fig{fig8} (circles) and it can be seen that 
for independent beads a
linear relation $\Delta \bar \eta \propto \Phi$
 holds  up to about $\Phi^\ast \simeq 0.015$. 
Beyond  $\Phi^\ast$ the effects of hydrodynamic particle-particle interaction becomes
increasingly stronger and this effect is described
in  \Eq{eq_ebg} by the  contribution  quadratic in $\Phi$. The simulation data
for a suspension of independent spheres  in \fig{fig8}  (circles) 
are approximated by the solid line due to  Eq.~(\ref{eq_ebg}) 
reasonably well up to  a volume fraction  $\Phi \approx 0.08$.

The mean distance $l$ between the 
centers of two spheres
decreases with the volume fraction  $\Phi$ as follows,
\begin{align}
 \label{l_eq}
 l=a\left( \frac{4\pi}{3\Phi} \right)^{1/3} \,.
\end{align}
At the volume fraction $\Phi^\ast$
one obtains a  mean distance $l^\ast \approx 6.5 a$ and for $l<l^\ast$
the hydrodynamic particle-particle interaction becomes
significant. In the case of dumbbells this length scale has to be compared
with the dumbbell extension $R_0$.

The contribution of a single dumbbell to $\Delta   \eta$ 
deviates from the contribution of two independent beads 
only significantly
during the short temporal range of a dumbbell turn 
in the range $\theta \sim \pi/2$, as indicated in 
\fig{fig5} and \fig{fig4a}.
Therefore, the temporally averaged dumbbell contribution to 
$\Delta \bar \eta$ 
is nearly equal to the contribution of  
two independent beads. 
However, this
deviation increases for a fixed number of dumbbells 
with the dumbbell extension $R_0$.

If one increases the volume fraction $\Phi$ of dumbbells,
their extension $R_0$ or both simultaneously, 
   their contribution
 to $\Delta \bar \eta$ 
deviates 
from the contribution of
the same number of independent beads because of the
following  trends. 
An increasing number  of dumbbells
increases the dumbbell turns per unit time
and therefore their dumbbell contribution to $\Delta \bar \eta$. 
In a dumbbell suspension
the motion of pairs of beads
is strongly correlated. This leads to an effective
extension of the hydrodynamic and excluded volume bead-bead interaction length
and simultaneously to a stronger enhancement of $\Delta \bar \eta$
for dumbbells with  $\Phi$ than for independent beads.
These effects together  cause the increasing deviations
between the data shown in \fig{fig8} 
for independent beads (circles) and
for  dumbbells of length $R_0=3.33a$ (triangles).

For longer dumbbells with $R_0=6.66a$
the hydrodynamic and excluded volume interaction is even more
enhanced. In addition the length $l$ becomes already
for  volume fractions $\Phi > 0.015$ smaller than $R_0$ 
 and the ranges motion of individual dumbbells 
start of overlap more and more with
increasing values of $\Phi$. Both trends
cause for dumbbells of length $R_0=6.66a$
an even larger contribution to $\Delta \bar \eta$
as for same density of dumbbells of length $R_0=3.33a$,
as can be seen by comparing in \fig{fig8}
the data obtained for 
dumbbells of length  $R_0=6.66a$ (squares) with those obtained for
dumbbells of length $R_0=3.33a$ (triangles).

\begin{figure}[ht]
\vspace{2mm}
  \begin{center}
  \includegraphics[width=0.9\columnwidth]{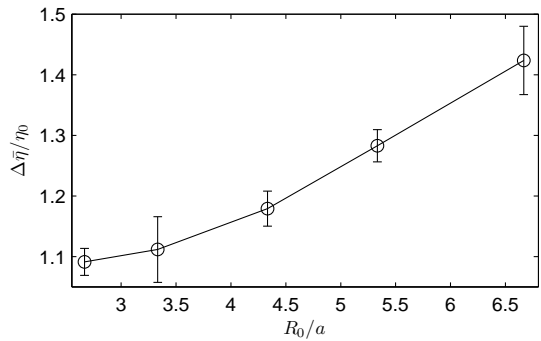}
  \end{center}
  \caption{The relative viscosity $\Delta \bar \eta$
 as a function of the relative length $R_0/a$ of the
 undistorted springs of the dumbbells in suspension which occupy
the volume fraction $\Phi=0.1056$ at $\Wi=0.33$. 
$\eta_0$ is the viscosity of a suspension of spheres at the same volume fraction.}
\label{fig9}
\end{figure}

How the dumbbell contribution to
  $\Delta \bar \eta$ increases as a function of
 the relative extension $R_0/a$
for a fixed dumbbell density   $\Phi=0.1056$, 
is shown in  \fig{fig9}. In this example  the effect of dumbbell-dumbbell 
interaction may enhance the shear viscosity in the range $R_0/a>6$
by more than 40 percent compared to the same volume fraction of
independent beads.

\section{Conclusion}

The dynamics of dumbbells in a linear shear flow
was investigated by fluid particle dynamics simulations.
The numerical results on the tumbling motion of a 
single dumbbell illustrate, 
that the effective viscosity in a shear cell is enhanced 
only during a short part of a dumbbell turn, compared to
the case of two independent beads in the fluid.
This enhancement of the
shear viscosity during a dumbbell turn, however, decreases
with increasing values of the Weissenberg number, as one
can recognize  by comparing the results shown in \fig{fig4} (a) and in \fig{fig4}(b).
This trend is similar as   shear thinning  in polymer
solutions.

The dumbbell contribution to the shear stress determined
via the Kramers-Kirkwood formula becomes negative  with
larger values of the Weissenberg number
for an increasing part of a dumbbells turn.
This negative contribution   increases with the ratio between the
bead diameters and the dumbbell length.
The dumbbell dynamics obtained in simultions
is often used to determine the dumbbell contribution to
the shear stress tensor via Kramers-Kirkwood formula.
 To the best of our knowledge, there was no comparison 
between a direction determination of the shear stress contribution of dumbbells
 and  calculations via  the Kramers-Kirk-wood formula.
It illustrates the limitation of the often used 
Kramers-Kirk-wood formula for a determination of the  dumbbell contribution 
 to the shear viscosity of a suspension
because we find in direct simulations at any stage of 
a dumbbell turn a positive contribution to the shear stress.

Up to a volume fraction $\Phi \simeq 0.08$
of independent spheres in a fluid, we found 
good agreement between the fluid particle simulations and 
the Batchelor-Green formula \cite{Batchelor:1972.1}
 for the effective  viscosity of a suspension of unconnected  spheres.
In the diluted regime  dumbbells cause only a slight enhancement
of the time-averaged shear viscosity due to 
 the following reasons.  
Since dumbbells are most of the  time nearly parallel
oriented to the undisturbed stream lines  they contribute  
to the time-averaged shear viscosity only slightly more than
  unconnected beads. In addition, 
in the diluted regime also the hydrodynamic and excluded volume
 dumbbell-dumbbell interaction is still small.

By increasing the volume fraction of dumbbells in
a fluid the hydrodynamic and excluded
volume interaction become increasingly more important than  for 
unconnected beads: In a dumbbell suspension a pair of beads
performs a correlated motion and therefore,
the effective hydrodynamic and excluded volume
interaction length is enhanced compared
to a suspension of independent beads. This leads in the case of 
a dumbbell suspension to a significantly stronger
contribution to the shear viscosity than
for unconnected beads. This effect is especially  enhanced when
the dumbbell extension exceeds the mean distance between the beads
of dumbbells.

Due to dumbbell-dumbbell interactions, the
 dynamics of a single dumbbell out of a suspension
is by far more complex than that of an
isolated one in shear flow, 
similar as the complex polymer dynamics in  elastic turbulence. 
The statistics of the complex dynamics of a suspension of 
dumbbells and trimers, the possibility of
shear thinning effects occurring for both types of bead-spring models in
 suspension as well as the possibility of
turbulent behavior in such suspensions is discussed 
elsewhere.

{\it {Acknowledgments.-}}
We would like to thank Chaouqi Misbah for useful discussions. This work was supported
by the by the Bayerisch-Franz\"osisches Hochschulzentrum and by the German science
foundation (DFG) through the priority program on micro- and nanofluidics SPP1164 and
through the research unit FOR608.

\bibliographystyle{prsty}

\end{document}